\documentclass[prl,reprint,superscriptaddress]{revtex4-1}

\usepackage{amsmath,graphicx,textcomp}
\usepackage{import}
\usepackage{siunitx}

\makeatletter
\renewcommand{\@seccntformat}[1]{}
\makeatother

\begin{document}


\title{Measuring Many-Body Distribution Functions in Fluids using Test-Particle Insertion}



\author{Adam Edward \surname{Stones}}
\email{adam.stones@chem.ox.ac.uk}

\affiliation{Department of Chemistry, Physical \& Theoretical Chemistry Laboratory, University of Oxford, Oxford OX1 3QZ, United Kingdom}
\author{Dirk G. A. L. \surname{Aarts}}
\affiliation{Department of Chemistry, Physical \& Theoretical Chemistry Laboratory, University of Oxford, Oxford OX1 3QZ, United Kingdom}


\date{\today}

\begin{abstract}
\noindent We derive a hierarchy of equations which allow a general $n$-body distribution function to be measured by test-particle insertion of between $1$ and $n$ particles, and successfully apply it to measure the pair and three-body distribution functions in a simple fluid. The insertion-based methods overcome the drawbacks of the conventional distance-histogram approach, offering enhanced structural resolution and a more straightforward normalisation. They will be especially useful in characterising the structure of inhomogeneous fluids and investigating closure approximations in liquid state theory.
\end{abstract}



\maketitle 

\noindent The distribution functions $g^{(n)}$ are central to the statistical mechanical description of classical fluids, characterising the short-ranged order of their particles' positions \cite{Hansen2013}. It is widely appreciated that knowledge of the pair distribution function $g^{(2)}$, where $n=2$, provides access to the thermodynamics of a fluid by integration, yielding the compressibility and, for pairwise-additive systems, the pressure and energy \cite{Hansen2013,Barker1976,Allen2017}. While this is correct, even simple monatomic liquids such as argon cannot be described adequately with pairwise interactions alone \cite{Axilrod1943,Barker1968,Barker1969,Barker1971,Present1971}, and fully incorporating many-body interactions requires knowledge of higher-order distribution functions, where $n\ge3$ \cite{Raveche1978}. Moreover, theoretical calculations of $g^{(2)}$ require consideration of many-body correlations even in the additive case \cite{Kirkwood1935,Haymet1981b}, and expressions for higher-order distribution functions in terms of lower ones can be used to close the Yvon-Born-Green (YBG) hierarchy in integral equation theories \cite{Henderson1967,Lee1968,Ree1971,Uehara1979a,Uehara1979b,Haymet1981a,Haymet1981b,Taylor1992}. Properties depending on thermodynamic derivatives, such as the heat capacity, are expressed as integrals over many-body distribution functions \cite{Schofield1966,Raveche1978,Haymet1981b}, as is the configurational entropy \cite{Nettleton1958,Raveche1971,Baranyai1989}. Finally, a description of fluid structure at the pairwise level alone is inherently limited: simulations \cite{Raveche1974,Raveche1978} and colloidal studies \cite{Russ2003,Zahn2003,Ho2006} have found that the triplet distribution function $g^{(3)}$ shows clear signatures of crystal-like order even in the fluid phase, at lower densities than $g^{(2)}$ does. Disruption of this incipient order by impurities \cite{Ho2006} or locally-preferred five-fold structures \cite{Frank1952,Steinhardt1983,Spaepen2000,Leocmach2012,Taffs2016} may play an important role in glass formation.

Previous measurements of many-body distribution functions using particle coordinates from simulations \cite{Alder1964,Rahman1964,Krumhansl1972,Wang1972,Raveche1972b,Raveche1974,Block1975,Tanaka1975,McNeil1983} and colloidal experiments \cite{Russ2003,Zahn2003,Ho2006} have used the conventional distance-histogram approach. While this method usually provides adequate resolution for $g^{(2)}$ in homogeneous fluids, which depends only on the interparticle separation $r$, for many-body distribution functions its shortcomings are increasingly pronounced. Even in the homogeneous case, $g^{(3)}$ depends on three variables, and doubling the spatial resolution reduces the bin volumes by a factor of 8, significantly increasing the statistical noise. For configurations where the particles approach closely, which are especially important in evaluating closure approximations \cite{Raveche1972a,Raveche1972b,Abramo1972,Block1975,Tanaka1975,Raveche1978,McNeil1983,Attard1991}, the distribution functions vary rapidly and approximate interpolations must be performed---in this work, we find both linear and cubic interpolations to be inadequate for $g^{(3)}$. Normalisation of the histograms is also increasingly complex: e.g.\ for $g^{(3)}$, the interparticle separations must satisfy triangle inequalities, which intersect the bins used in a naive distance-histogram scheme \cite{Krumhansl1972,Tanaka1975}.

In this Letter, we address these problems by deriving a hierarchy of equations for measuring $g^{(n)}$ by test-particle insertion \cite{Widom1963,Stones2018}. The hierarchy allows a given $g^{(n)}$ to be obtained by multiple routes, corresponding to insertion of between $1$ and $n$ test particles. For insertion of $n$ or $n\!-\!1$ test particles, the measurements are exact: $g^{(n)}$ can be measured with arbitrary resolution and for specific particle configurations. Measurements based on inserting fewer test particles require a histogram in fewer variables than the full distance-histogram method, considerably reducing the statistical noise, while the normalisation of the insertion measurements is straightforward in all cases. We demonstrate the application of the hierarchy to measure $g^{(2)}$ and $g^{(3)}$ in a simple homogeneous fluid, comparing the results obtained using different numbers of insertions with each other and to those of the distance-histogram method. In particular, we find that the insertion-based methods remove the need to interpolate $g^{(3)}$ and are therefore particularly advantageous when examining closure approximations.

The $n$-body density in the grand ensemble is written as \cite{Hansen2013}

\begin{widetext}
\begin{equation}
   \rho^{(n)}(\mathbf{s}^n) = \frac{1}{\Xi} \sum_{N=n}^{\infty} \frac{z^N}{(N-n)!} \int \mathrm{d}\mathbf{r}^{(N-n)} \exp\left[-\beta U_N(\mathbf{s}^n, \mathbf{r}^{(N-n)})\right]\mathrm{,}
\label{eq:nDensity}
\end{equation}
\end{widetext}
where $U_N$ is the potential energy of $N$ interacting particles, which includes interactions with external fields, the inverse temperature $\beta=1/k_{\mathrm{B}}T$, and $z$ is the reduced activity $\exp(\beta\mu)/\Lambda^d$, with $\mu$ the chemical potential, $\Lambda$ the thermal wavelength, and $d$ the system dimensionality. The grand partition function $\Xi$ is given by
\begin{equation}
   \Xi = \sum_{N=0}^{\infty} \frac{z^N}{N!} \int \mathrm{d}\mathbf{r}^{N} \exp\left[-\beta U_N(\mathbf{r}^N)\right]\mathrm{.}\nonumber
\label{eq:GPF}
\end{equation}
We use $\mathbf{r}_i$ to denote a coordinate that varies under integration and $\mathbf{s}_i$ to denote a fixed position in space, and adopt the notational conventions $\mathbf{r}^N\equiv\mathbf{r}_1, \dots, \mathbf{r}_N$ and $\mathbf{r}^{(N-n)}\equiv\mathbf{r}_{n+1}, \dots, \mathbf{r}_N$ \cite{Hansen2013}. Inspecting Eq.\ (\ref{eq:nDensity}), we see that $\rho^{(n)}$ is the ratio of the sum of all microstates with any $n$ particles at $\mathbf{s}^n$ to the total sum of microstates, with the microstates weighted by their probability densities in the grand ensemble. The $n$-body density is therefore the marginal probability density of having any $n$ particles at the positions $\mathbf{s}^n$.

The conditional probability density $\rho^{(n-m)}(\mathbf{s}^{(n-m)}|\mathbf{s}^m)$ of having $(n-m)$ particles at $\mathbf{s}^{(n-m)}$ given that there are $m$ particles at $\mathbf{s}^m$ is therefore given by the ratio of $\rho^{(n)}(\mathbf{s}^n)$ to $\rho^{(m)}(\mathbf{s}^m)$ \cite{Percus1962,Percus1964,Puoskari2001}. Using Eq.\ (\ref{eq:nDensity}),

\begin{widetext}
\begin{equation}
   \frac{\rho^{(n)}(\mathbf{s}^n)}{\rho^{(m)}(\mathbf{s}^m)} = \frac{\sum_{N=n}^{\infty} \frac{z^N}{(N-n)!} \int \mathrm{d}\mathbf{r}^{(N-n)} \exp\left[-\beta U_N(\mathbf{s}^n, \mathbf{r}^{(N-n)})\right]}{\sum_{N=m}^{\infty} \frac{z^N}{(N-m)!} \int \mathrm{d}\mathbf{r}^{(N-m)} \exp\left[-\beta U_N(\mathbf{s}^m, \mathbf{r}^{(N-m)})\right]}\mathrm{.}
\label{eq:nmDensityRatio}
\end{equation}
\end{widetext}
Extending the arguments of Widom \cite{Widom1963,Rowlinson1984,Lee1995,Robinson2019}, we split $U_N$ in the numerator of Eq.\ (\ref{eq:nmDensityRatio}) into a term representing all of the interactions between particles at $\mathbf{s}^m$ and $\mathbf{r}^{(N-n)}$, and the additional potential energy $\Psi(\mathbf{s}^{(n-m)};\mathbf{s}^m,\mathbf{r}^{(N-n)})$ arising from interactions of the particles at $\mathbf{s}^{(n-m)}$ with each other and the other particles of the system. The numerator becomes
\begin{widetext}
\begin{align}
      & \sum_{N=n}^{\infty} \frac{z^N}{(N-n)!} \int \mathrm{d}\mathbf{r}^{(N-n)} \exp\left[-\beta U_{N-(n-m)}(\mathbf{s}^m, \mathbf{r}^{(N-n)})\right] \exp\left[-\beta \Psi(\mathbf{s}^{(n-m)};\mathbf{s}^m,\mathbf{r}^{(N-n)})\right]\mathrm{,}\nonumber\\
    = & z^{n-m}\sum_{M=m}^{\infty} \frac{z^M}{(M-m)!} \int \mathrm{d}\mathbf{r}^{(M-m)} \exp\left[-\beta U_{M}(\mathbf{s}^m, \mathbf{r}^{(M-m)})\right] \exp\left[-\beta \Psi(\mathbf{s}^{(n-m)};\mathbf{s}^m,\mathbf{r}^{(M-m)})\right]\mathrm{,}
\label{eq:numerator}
\end{align}
\end{widetext}
where on the second line we have made the substitution $M=N-(n-m)$ and renumbered $\mathbf{r}^{(N-n)}$ as $\mathbf{r}^{(M-m)}$. When this is substituted into Eq.\ (\ref{eq:nmDensityRatio}), the right-hand side has the form of an ensemble average of $\exp(-\beta\Psi)$ over the states with particles at $\mathbf{s}^m$, multiplied by $z^{n-m}$. Hence,
\begin{align}
   \frac{\rho^{(n)}(\mathbf{s}^n)}{\rho^{(m)}(\mathbf{s}^m)} &= z^{n-m} \left\langle \exp[-\beta \Psi(\mathbf{s}^{(n-m)};\mathbf{s}^m,\mathbf{r}^{(N-m)})] \right\rangle_{\mathbf{s}^m}\mathrm{,}\nonumber 
\\
                                                             &= z^{n-m} P^{(n-m)}(\mathbf{s}^{(n-m)}|\mathbf{s}^m)\mathrm{,}
\label{eq:densityRelationInsertion}
\end{align}
where on the final line we have simplified the notation by using $P$ for the ensemble average.

Introducing the definition of the distribution functions \cite{Hansen2013}
\begin{equation}
   g^{(n)}(\mathbf{s}^n) = \frac{\rho^{(n)}(\mathbf{s}^n)}{\prod_{i=1}^{n} \rho^{(1)}(\mathbf{s}_i)}
\label{eq:DistributionFunctions}
\end{equation}
and rearranging Eq.\ (\ref{eq:densityRelationInsertion}), we obtain
\begin{equation}
   g^{(n)}(\mathbf{s}^{n}) = z^{n-m} g^{(m)}(\mathbf{s}^{m}) \frac{P^{(n-m)}(\mathbf{s}^{(n-m)}|\mathbf{s}^m)}{\prod_{i=m+1}^{n} \rho^{(1)}(\mathbf{s}_i)}\mathrm{.}
\label{eq:Widom}
\end{equation}
Finally, we use $\rho^{(1)}(\mathbf{s}_i) = z P^{(1)}(\mathbf{s}_i)$, which may be obtained by setting $n=1$ and $m=0$ in Eq.\ (\ref{eq:densityRelationInsertion}), and is simply Widom's result for inhomogeneous systems \cite{Widom1978}, yielding
\begin{equation}
   g^{(n)}(\mathbf{s}^{n}) = g^{(m)}(\mathbf{s}^{m}) \frac{P^{(n-m)}(\mathbf{s}^{(n-m)}|\mathbf{s}^m)}{\prod_{i=m+1}^{n} P^{(1)}(\mathbf{s}_i)}\mathrm{.}
\label{eq:central}
\end{equation}

\begin{figure}[!ht]
    \includegraphics[width=\columnwidth]{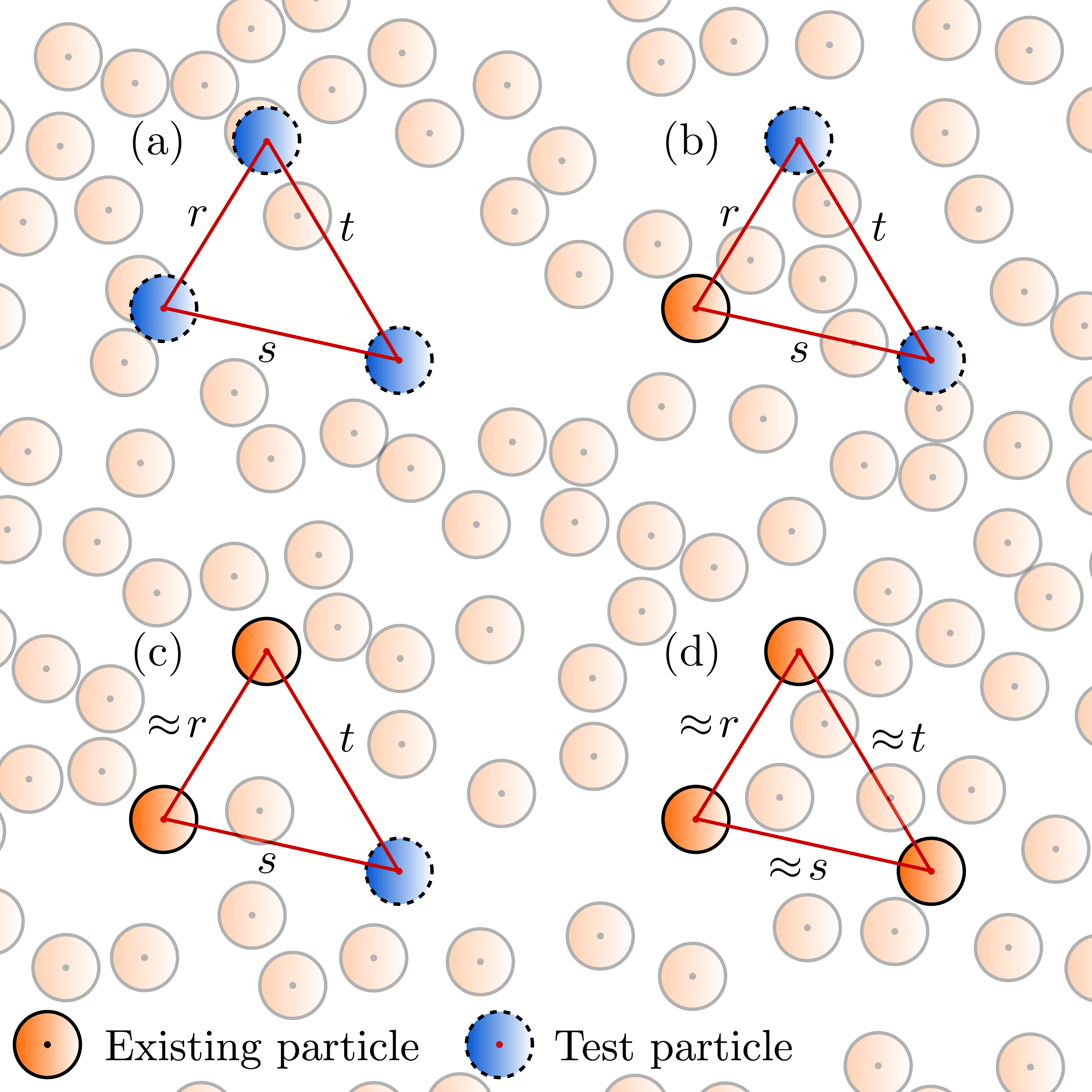}
    \caption{Applying Eq.\ (\ref{eq:central}) permits $g^{(3)}(r,s,t)$ to be measured by insertion of up to three test particles, (a)-(c), in addition to the conventional distance-histogram approach, (d). Methods (a) and (b) are exact, while (c) and (d) require histograms in one and three variables respectively.}
    \label{fig:schematic}
\end{figure}

This hierarchy of equations is the central result of this letter, and allows an $n$-body distribution function to be expressed as a product of a lower-order $m$-body distribution function and a ratio of ensemble averages which may be evaluated by test-particle insertion. Here, we will illustrate its use by applying it to measure $g^{(2)}$ and $g^{(3)}$ in a homogeneous system, where the equations are particularly straightforward. We stress, however, that Eq.\  (\ref{eq:central}) is general and can in principle be used to measure distribution functions of arbitrary order $n$ in inhomogeneous systems.


Substituting $n=2$ into Eq.\ (\ref{eq:central}) gives
\begin{subequations}
\begin{align}
   g^{(2)}(\mathbf{s}_1,\mathbf{s}_2) & = \frac{P^{(2)}(\mathbf{s}_1,\mathbf{s}_2)}{P^{(1)}(\mathbf{s}_1)P^{(1)}(\mathbf{s}_2)}\mathrm{,} & m=0\mathrm{,}\label{eq:inhom_g2_2}\\
                                      & = \frac{P^{(1)}(\mathbf{s}_2|\mathbf{s}_1)}{P^{(1)}(\mathbf{s}_2)}\mathrm{,}                      & m=1\mathrm{,}\label{eq:inhom_g2_1}
\end{align}
\end{subequations}
where we have noted that $g^{(0)}=g^{(1)}=1$. In a homogeneous system, $P^{(1)}(\mathbf{s}_i)\equiv P$ is spatially uniform, while $g^{(2)}(\mathbf{s}_1,\mathbf{s}_2)$, $P^{(2)}(\mathbf{s}_1,\mathbf{s}_2)$ and $P^{(1)}(\mathbf{s}_2|\mathbf{s}_1)$ depend only on the separation $|\mathbf{s}_2-\mathbf{s}_1|=r$, such that
\begin{subequations}
\begin{align}
   g^{(2)}(r) & = \frac{P^{(2)}(r)}{P^2}\mathrm{,}              & m=0\mathrm{,}\label{eq:hom_g2_2}\\
              & = \frac{P^{(1)}(r)}{P}\mathrm{.}     & m=1\mathrm{.}\label{eq:hom_g2_1}
\end{align}
\end{subequations}

There are therefore two possibilities for measuring $g^{(2)}$ by test-particle insertion. According to Eq.\ (\ref{eq:hom_g2_2}), we first evaluate $P^{(2)}(r)$ by test-insertions of a pair of particles with separation $r$ at random positions in the fluid; dividing by the square of $P$, which can be measured by one-particle insertions, then yields $g^{(2)}$. Alternatively, we can perform test-insertions of one particle at a fixed separation $r$ from an existing particle of the fluid; this yields $P^{(1)}(r)$ which can be used in Eq.\ (\ref{eq:hom_g2_1}) \footnote{Note that to preserve the simplicity of the notation, the conditional nature of $P^{(1)}(r)$, $P^{(2)}(r,s,t)$ and $P^{(1)}(r,s,t)$ is taken to be implicit.}. In contrast with the distance-histogram method, both insertion methods are exact and $g^{(2)}$ can be obtained at arbitrary resolution \cite{Stones2018}.

In a homogeneous fluid, $g^{(3)}$ is a function of the triangle formed with a particle at each vertex, which here is specified by the side lengths $r$, $s$ and $t$, as shown in Fig.\ \ref{fig:schematic}. In this case, Eq.\ (\ref{eq:central}) gives
\begin{subequations}
\begin{align}
   g^{(3)}(r,s,t) & = \frac{P^{(3)}(r,s,t)}{P^3}\mathrm{,}           & m=0\mathrm{,}\label{eq:hom_g3_3}\\
                  & = \frac{P^{(2)}(r,s,t)}{P^2}\mathrm{,}           & m=1\mathrm{,}\label{eq:hom_g3_2}\\
                  & = \frac{P^{(1)}(r,s,t)}{P}g^{(2)}(r)\mathrm{,}   & m=2\mathrm{,}\label{eq:hom_g3_1}
\end{align}
\end{subequations}
with the corresponding insertion methods summarised in Fig.\ \ref{fig:schematic}(a)-\ref{fig:schematic}(c). In the first case, we evaluate $P^{(3)}(r,s,t)$ by test-insertions of a triangle of three particles at a random position and orientation in the fluid [Fig.\ \ref{fig:schematic}(a)]. In the second case, we measure $P^{(2)}(r,s,t)$ using two-particle insertions with an existing particle as the third vertex [Fig.\ \ref{fig:schematic}(b)]; while in the third case we obtain $P^{(1)}(r,s,t)$ by test-insertions of one particle around pairs of existing particles separated by one of the side lengths \cite{Note1} [Fig.\ \ref{fig:schematic}(c)].

Note that in this final case, we must also multiply by $g^{(2)}(r)$, where $r$ is the separation of the pair of particles already in the system. Since $g^{(2)}(r)$ may be also be measured by one-particle insertions, this illustrates the principle that Eq.\ (\ref{eq:central}) can be used to express the $g^{(n)}$ as products of ensemble averages corresponding only to one-particle insertions, analogous to the \emph{bootstrap} operation described by Percus \cite{Percus1964}. The first two cases are again exact and allow $g^{(3)}$ to be measured precisely for specific triangles, while the third requires a histogram in the side length $r$ only, significantly improving the statistics compared with the full distance-histogram approach. In all cases, normalisation is just a straightforward division by the appropriate power of $P$.

We next test these methods using two-dimensional Monte Carlo simulations in the grand ensemble, with the particles interacting via a purely-repulsive Weeks-Chandler-Andersen (WCA) pair potential \cite{Weeks1971}. We performed three simulations, with reduced densities $\rho\sigma^2 \approx 0.41$, $\rho\sigma^2 \approx 0.60$ and $\rho\sigma^2 \approx 0.77$; here, we focus on the lowest density to illustrate the formal equivalence of the insertion methods. The results for the higher densities, where the two- and three-particle insertion methods are more noisy, may be found in the Supplemental Material, along with further details of the simulations \cite{supplementalMaterial}.

\begin{figure*}[!ht]
    \includegraphics[width=\textwidth]{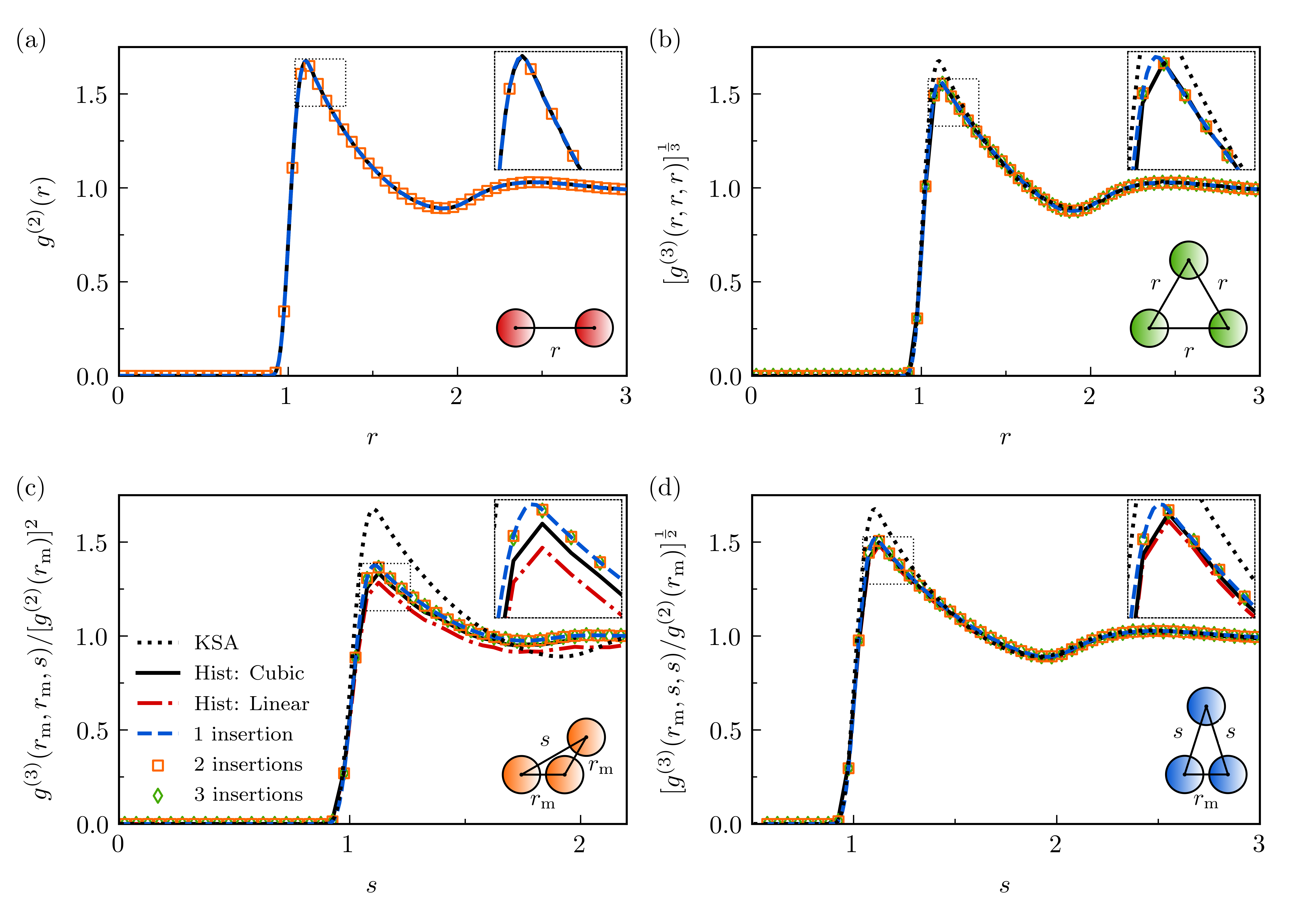}
    \caption{Comparison of $g^{(2)}$, (a), and $g^{(3)}$ for equilateral and isosceles triangles, (b)-(d), measured by the distance-histogram method and each of the insertion methods, in the fluid with $\rho\sigma^2 \approx 0.41$. The insets show a zoom of each plot around the first maximum. The ordinates of (b)-(d) facilitate comparison with $g^{(2)}$, which is the KSA prediction. In (c) and (d), the side length $r_{\mathrm{m}}$, corresponding to the first maximum of $g^{(2)}$, was not at the centre of one of the bins used in the distance-histogram method---instead, the results of linear and cubic interpolation are shown \cite{captionRef}. See the Supplemental Material for more details \cite{supplementalMaterial}.}
    \label{fig:configurations}
\end{figure*}

We show the results for $g^{(2)}$ in Fig.\ \ref{fig:configurations}(a). The agreement between the results of the one-particle insertion and distance-histogram methods is perfect, as previously found for hard disks away from contact \cite{Stones2018}. Here we demonstrate that the agreement also extends to continuous interactions, and that these results agree with those of the two-particle insertion method. 

Since $g^{(3)}$ is a function of three variables, it is more difficult to visualise than $g^{(2)}$. Typically, only a subset of $g^{(3)}$ is shown, either where the side lengths are functions of a single variable, e.g.\ for equilateral and isosceles triangles \cite{Alder1964,Rahman1964,Krumhansl1972,Wang1972,Raveche1974,Block1975,Tanaka1975,McNeil1983}, or where two particles are held at a fixed separation and $g^{(3)}$ is plotted as function of the third \cite{Raveche1972b,Zahn2003,Russ2003,Ho2006}. The insertion approach is efficient in both cases, since $g^{(3)}$ can be measured directly for the subset of triangles without having to consider all triplets of particles in the system. Here we opt for the first case, showing $g^{(3)}$ measured for equilateral and isosceles triangles in Figs.\ \ref{fig:configurations}(b)-\ref{fig:configurations}(d).

For the equilateral triangles [Fig.\ \ref{fig:configurations}(b)], the agreement of the insertion results with each other and those of the distance-histogram method is essentially perfect, verifying the validity of Eq.\ (\ref{eq:central}) and the insertion methods described above. The two- and three-particle insertion methods are formally exact, while the one-particle insertion method has a resolution comparable to the distance-histogram measurement of $g^{(2)}$, since a histogram is only required in the separation $r$ of the existing particles [Fig.\ \ref{fig:schematic}(c)]. The implementation of the one-particle insertion method is highly efficient: we generally attempt fewer insertions than with the other insertion methods, with attempts only made around existing pairs of particles which have approximately the correct separation. Although this does not allow for measurements when $r$ is inside the core region, since there are no existing pairs with these separations, this limitation has no practical consequence---at such separations, $g^{(2)}=0$ and so $g^{(3)}=0$ according to Eq.\ (\ref{eq:hom_g3_1}) \footnote{For continuous pair potentials such as that used here, $g^{(2)}$ in the core region is actually finite but extremely small, and so is measured as zero in practice.}. 

Figures \ref{fig:configurations}(c) and \ref{fig:configurations}(d) show $g^{(3)}$ measured for isosceles triangles, with either one or two side lengths fixed as $r_{\mathrm{m}}$, corresponding to the first maximum in $g^{(2)}$. Note that the range of values for the other side length $s$ is restricted by the triangle inequality. The agreement of the insertion measurements with each other is again excellent, but these results highlight a key deficiency of the distance-histogram approach. To obtain adequate statistics, larger bins are required when measuring $g^{(3)}$ than when measuring $g^{(2)}$; consequently, $r_{\mathrm{m}}$ does not lie at the centre of a $g^{(3)}$ bin, and an approximate interpolation must be performed. Although a cubic interpolation \cite{Walker2019} outperforms a simple linear scheme, there is still a marked deviation from the insertion results, especially in Fig.\ \ref{fig:configurations}(c). By contrast, no issues with resolution arise for the two-particle and three-particle insertion methods, which are formally exact, while the one-particle method again allows for a much higher resolution than the pure distance-histogram approach. The poor resolution of the distance-histogram method is expected to be still more pronounced when measuring higher-order distribution functions, where the dependence on more variables further compromises the resolution and makes interpolation schemes more challenging to implement.

For each plot, the ordinate is chosen to facilitate comparison with $g^{(2)}$, which is the prediction of the Kirkwood Superposition Approximation (KSA) closure \cite{Barker1976}
\begin{equation}
   g^{(3)}(r,s,t) = g^{(2)}(r)g^{(2)}(s)g^{(2)}(t)\mathrm{,}
   \label{eq:KSA}
\end{equation}
which assumes that the correlation of any two particles is not affected by the presence of the third \cite{Attard1989}. The greatest deviations from the KSA are found when all three particles are close together; this is especially clear in Fig.\ \ref{fig:configurations}(c), where the isosceles triangles are analogous to rolling-contact configurations in hard-sphere systems \cite{Uehara1979b,Attard1991,Attard1992,Muller1993,Kalyuzhnyi2019}, and where the poor resolution of the distance-histogram method is also most evident. Hence, while comparisons with more sophisticated closures of the YBG hierarchy are possible \cite{Raveche1972a,Raveche1972b,Abramo1972,Block1975,Tanaka1975,Raveche1978,McNeil1983}, comparing with the KSA already highlights the advantages of the insertion approach in examining such approximations.

Which insertion method should be used? The one-particle insertion method is particularly efficient and offers excellent resolution when measuring $g^{(3)}$, with the need for a histogram in only one variable overcoming the poor resolution of the distance-histogram method. On the other hand, the remaining histogram may lead to increased noise in lower-density systems, or when measuring higher-order $g^{(n)}$, where a histogram in more than one variable is again required. Although the insertion method with insertion of $n$ or $n\!-\!1$ particles is formally exact, multi-particle insertions exacerbate the well-documented breakdown of the test-particle insertion approach at high densities \cite{Kofke1997,supplementalMaterial}, while approaches based on one-particle insertions are expected to work well except in very high-density fluids and solids \cite{Stones2018}. The optimum number of insertions therefore depends on the order $n$ of the distribution function, the state of the fluid, and the application in mind.

\emph{Conclusion}---We have derived a hierarchy of insertion-based methods for measuring many-body distribution functions and demonstrated their application in measuring $g^{(2)}$ and $g^{(3)}$. The methods address drawbacks of the conventional distance-histogram approach, offering improved resolution and a more straightforward normalisation. They are expected to be particularly advantageous in inhomogeneous systems and when scrutinising various closure approximations used in integral equation theories. As well as being effective in measuring $g^{(3)}$, they will facilitate investigations into expressions for higher-order distribution functions, such as $g^{(4)}$, which can be used to close the YBG hierarchy at higher levels \cite{Lee1968,Ree1971,Uehara1979a,Uehara1979b,Raveche1978}.

\begin{acknowledgments}
The authors would like to thank Roel Dullens for critically reading the manuscript. A.E.S. gratefully acknowledges financial support from the University of Oxford Clarendon Fund.
\end{acknowledgments}


\bibliography{many_body_distribution}

\begin{thebibliography}{62}
\bibitem [{cap()}]{captionRef}%
  \BibitemOpen
  \href@noop {} {}\bibinfo {note} {Note that no interpolation was required in
  the equilateral case [Fig.\ \ref{fig:configurations}(b)], since the side
  lengths \emph{do} correspond with the bin centres.}\BibitemShut {Stop}%
\end{thebibliography}

\end{document}



\title{Supplemental Material: Measuring Many-Body Distribution Functions in Fluids using Test-Particle Insertion}



\author{Adam Edward \surname{Stones}}
\email{adam.stones@chem.ox.ac.uk}

\affiliation{Department of Chemistry, Physical \& Theoretical Chemistry Laboratory, University of Oxford, Oxford OX1 3QZ, United Kingdom}
\author{Dirk G. A. L. \surname{Aarts}}
\affiliation{Department of Chemistry, Physical \& Theoretical Chemistry Laboratory, University of Oxford, Oxford OX1 3QZ, United Kingdom}


\date{\today}


\maketitle 

\section{Results at Higher Densities}

\begin{figure*}[!ht]
    \includegraphics[width=\textwidth]{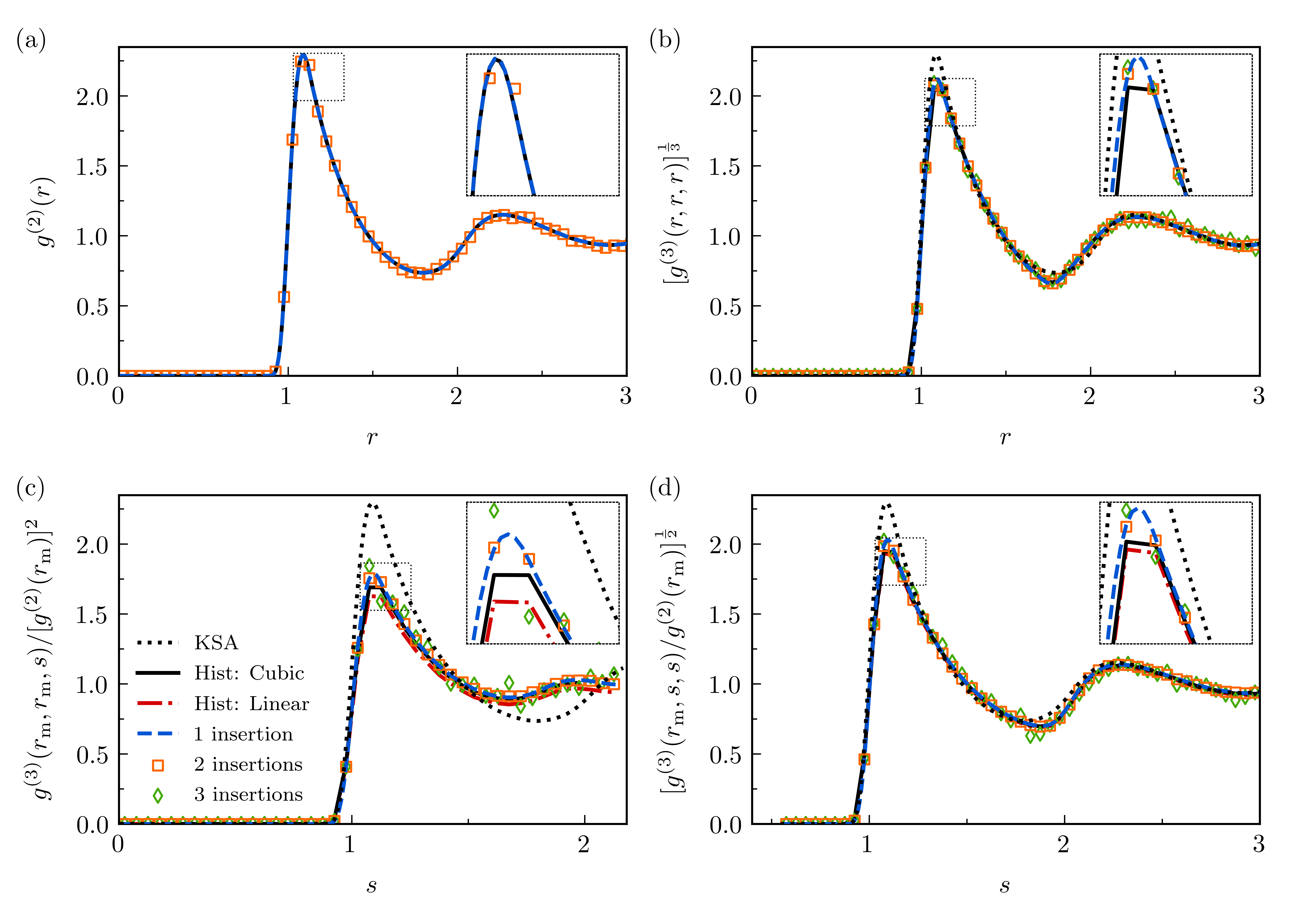}
    \caption{The same measurements shown in Fig.\ 2 of the main text, but performed for the system with $\rho\sigma^2 \approx 0.60$.}
    \label{fig:configurations1500}
\end{figure*}

\begin{figure*}[!ht]
    \includegraphics[width=\textwidth]{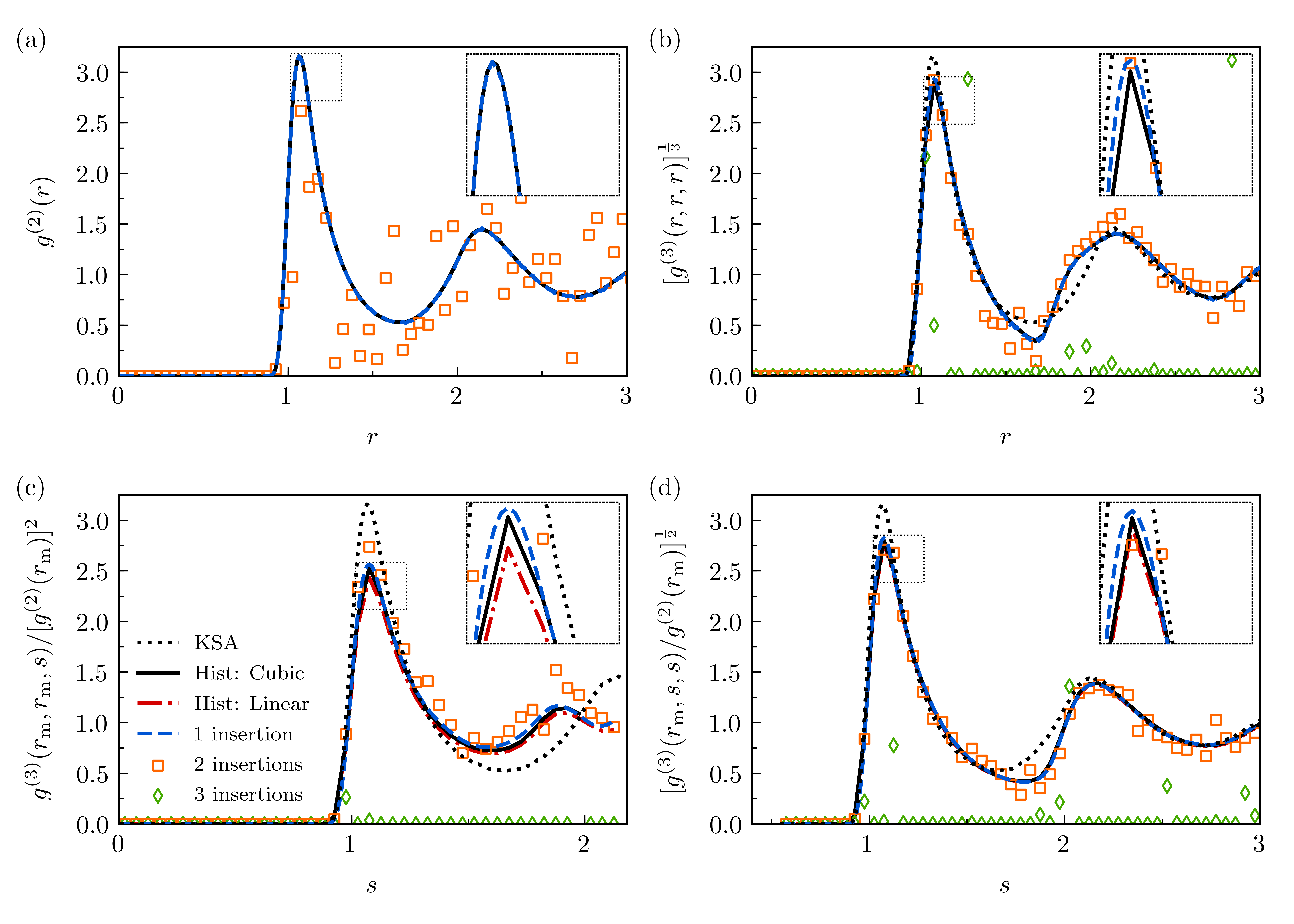}
    \caption{The same measurements shown in Fig.\ 2 of the main text, but performed for the system with $\rho\sigma^2 \approx 0.77$. The three-particle insertion method fails completely, and some points are outside of the plotted range.}
    \label{fig:configurations2000}
\end{figure*}

Figures \ref{fig:configurations1500} and \ref{fig:configurations2000} show the same measurements as Fig.\ 2 of the main text, but for the higher-density fluids with $\rho\sigma^2 \approx 0.6$ and $\rho\sigma^2 \approx 0.77$, respectively. As with the lower-density fluid, in both cases $g^{(3)}$ is best-captured by the one-particle insertion method, which we performed at a resolution $5\times$ that of the other measurements due to its efficient implementation and low noise. Once again, $r_{\mathrm{m}}$ did not lie at the centre of one of the bins used for the distance-histogram measurement of $g^{(3)}$, necessitating an approximate interpolation in the isosceles cases. As with the lower-density fluid, the deviation between the distance-histogram and one-particle insertion results is greatest in panel (c), where all three particles are close together at all allowed values of $s$ and the deviation remains even after cubic interpolation. Note also that the Kirkwood Superposition Approximation (KSA) does not perform as well as in the lower-density fluid, reflecting the increased importance of three-body correlations at higher densities: this is clearly evident in panels (b) and (d), which now show sizeable deviations at larger separations as well as at the first maximum. Finally, the equilateral $g^{(3)}$ at $\rho\sigma^2 \approx 0.77$ (Fig. \ref{fig:configurations2000}) shows a developing shoulder to the left of the second peak, indicating incipient crystalline order \cite{Ho2006}. 

The increased noise of the multi-particle insertion methods is more apparent at the higher densities, especially that of the three-particle insertion method which fails completely for $\rho\sigma^2 \approx 0.77$ (Fig.\ \ref{fig:configurations2000}). The two-particle insertion method also shows increased noise for both the $g^{(2)}$ and $g^{(3)}$ measurements. The breakdown of test-particle insertion methods at high densities is well-known in the context of chemical potential measurements \cite{Kofke1997}, and has the same origin here. In brief, it becomes too difficult to insert an additional particle: small or negative values of $\Psi$ are encountered only rarely, despite making an important contribution to the ensemble average. This difficulty is clearly greater still when inserting more particles, leading to a larger error in these insertion measurements and the breakdown of the multi-particle insertion methods at lower densities than in the one-particle case. At $\rho\sigma^2 \approx 0.77$, a small amount of noise is visible even in the one-particle case---while this may lead to the distance-histogram method being the method of choice for obtaining $g^{(2)}$, the superior resolution of the insertion route is still strongly apparent in the $g^{(3)}$ measurements (see the inset figures).


\section{Simulation}

\noindent We performed Monte Carlo in the grand ensemble. The interparticle interactions were pairwise according to a WCA potential \cite{Weeks1971}
\begin{equation}
   u_{\mathrm{WCA}}(r) = \left\{\begin{array}{l@{\quad}l}
    4\epsilon\left[\left(\frac{\sigma}{r}\right)^{12}-\left(\frac{\sigma}{r}\right)^6\right]+\epsilon\mathrm{,}%
    &\ r\le2^{\frac{1}{6}}\sigma\mathrm{,}\\
    0\mathrm{,}&\ r>2^{\frac{1}{6}}\sigma\mathrm{,}\\
    \end{array}\right.
\end{equation}
where $r$ is the interparticle separation. We used a square simulation box with side length $50\sigma$, and chose the interaction parameters $\sigma = 1$, $\epsilon=1$, and the inverse temperature $\beta=1/k_{\mathrm{B}}T=1$. We set the simulation chemical potential $$\mu^{\prime} = \mu-2k_{\mathrm{B}}T\ln(\Lambda/\sigma)$$ such that $\mu^{\prime}=1.25$, $\mu^{\prime}=4$ or $\mu^{\prime}=8.5$. These gave rise to mean numbers of particles $\overline{N} \approx 1017$ (reduced density $\rho\sigma^2 \approx 0.41$), $\overline{N} \approx 1507$ ($\rho\sigma^2 \approx 0.60$), and $\overline{N} \approx 1936$ ($\rho\sigma^2 \approx 0.77$), respectively.

We performed particle displacement, deletion and insertion moves in the ratio 3:1:1. For the displacements, the trial particle position was randomly chosen from a square box around the original position, with a maximum displacement in each direction of $0.6$ ($\rho\sigma^2 \approx 0.41$, acceptance ratio 52\%), $0.28$ ($\rho\sigma^2 \approx 0.60$, 51\%) and $0.19$ ($\rho\sigma^2 \approx 0.77$, 42\%). We initially placed ten particles at random in the box, and then allowed the system to equilibrate for $10^7$ moves, such that the equilibrium density was attained. We subsequently took snapshots every $10^4$ moves and saved them for analysis. The total number of analysed configurations was $5\times10^4$.

\section{Distance-Histogram Method}

For both the $g^{(2)}$ and $g^{(3)}$ measurements, distance criteria were used to avoid considering pairs and triangles of particles with large separations or side lengths, where the short-ranged structure of the fluid has essentially decayed away. For $g^{(2)}(r)$, pairs with a separation $r\le10\sigma$ were found and a histogram of the separations was made. This was normalised by the volumes of the annular bins and the number of particles considered to yield $\rho^{(1)}(r|\mathbf{0})$, the conditional one-body density given a particle at the origin $\mathbf{0}$. Finally, this was divided by $\rho$ to give $g^{(2)}(r)$.

For $g^{(3)}$, triangles were found with at least two sides $\le 5\sigma$; using the triangle inequality, the maximum length of the third side is then constrained to be $\le 10\sigma$. A 3D histogram of the side lengths was made with a bin-size of $0.05\sigma$ in each variable. In principle, the symmetry of $g^{(3)}$ means each triangle can be included six times (once for each permutation of $\{r,s,t\}$), though in practice, those triangles with the third side length $>5\sigma$ were only included twice due to the distance criteria. As noted in the introduction of the main text, care must be taken when normalising the histogram to only include the region of each bin which satisfies the triangle inequalities---we used a two-dimensional version \cite{Russ2003} of the procedure outlined in \cite{Krumhansl1972}. Normalisation by this volume and the number of particles considered gave $\rho^{(2)}(r,s,t|\mathbf{0})$, which was divided by $\rho^2$ to give $g^{(3)}(r,s,t)$.

In Figs.\ 2(a) we show the measured $g^{(2)}(r)$ at values of $r$ corresponding to the centre of each bin. In Fig.\ 2(b), we use $g^{(3)}$ bins with centres corresponding to equilateral configurations. As noted in the main text, finding $g^{(3)}$ is more challenging for the triangles used in Fig.\ 2(c) and Fig.\ 2(d) since $r_{\mathrm{m}}$, the separation at the primary maximum of the measured $g^{(2)}$ ($r_{\mathrm{m}}=1.105$ at $\rho\sigma^2 \approx 0.41$, $r_{\mathrm{m}}=1.085$ at $\rho\sigma^2 \approx 0.60$, and $r_{\mathrm{m}}=1.065$ at $\rho\sigma^2 \approx 0.77$), does not coincide with the centre of the bins used when measuring $g^{(3)}$, which have to be larger to provide adequate statistics. We therefore have to interpolate the measured function to provide values for comparison with the insertion methods and the KSA closure. This is particularly important in the case of Fig.\ 2(c), where the three particles are always close together and $g^{(3)}$ varies significantly on the length-scale of the bin size---a linear interpolation is clearly inadequate. While the cubic interpolation \cite{Walker2019} provides somewhat improved agreement with the insertion results, it comes with the disadvantage that it does not work when $s$ is too close to zero or the limits given by the triangle inequality, since in these cases the scheme requires values for bins where $g^{(3)}(r,s,t)$ is not defined.

\section{Test-particle Insertion Methods}

\textbf{$g^{(2)}$: Two-particle Test-Insertions.} We measured $g^{(2)}$ at separations corresponding to the centres of the bins used in the distance-histogram measurement, with a maximum $r$ of $5.025\sigma$. To measure $P^{(2)}(r)$, 1000 pairs of particles were inserted per configuration for each separation, at random locations and orientations. The inserted pairs were also separately considered as one-particle insertions to measure $P$, and Eq.\ (9a) was used to calculate $g^{(2)}(r)$.

\textbf{$g^{(2)}$: One-particle Test-Insertions.} The number of separations was $5\times$ that used in the two-particle insertion case, since this method provides better statistics. To measure $P^{(1)}(r)$, one insertion per particle was attempted for each separation in all configurations. In each case, the angle of the inserted particle relative to the existing particle was chosen at random from a uniform distribution. To measure $P$, a separate measurement was performed where $10^4$ test particles per configuration were inserted at random points in the fluid, and Eq.\ (9b) was used to calculate $g^{(2)}(r)$.

\textbf{$g^{(3)}$: Three-particle Test-Insertions.} Separations ($r$ or $s$) were chosen to correspond to the bin centres used in the distance-histogram measurement. To measure $P^{(3)}$, 1000 test-particle triangles were inserted for each separation in the configurations specified in the cartoons of Fig.\ 2(b)-2(d), at random locations and orientations. A separate measurement of $P$ was performed using $10^4$ test particles, and Eq.\ (10a) was used to calculate $g^{(3)}$.

\textbf{$g^{(3)}$: Two-particle Test-Insertions.} The same separations were used as for the three-particle insertion case. For each separation, two particles were inserted around each particle of the fluid in all configurations, to form a triangle with the appropriate side lengths [see Fig.\ 1(b)]. The triangles were constructed with a random orientation, and the insertions were used to measure $P^{(2)}$. A separate measurement of $P$ was performed using $10^4$ test particles, and Eq.\ (10b) was used to calculate $g^{(3)}$.

\textbf{$g^{(3)}$: One-particle Test-Insertions.} In the equilateral case, all pairs of particles with $r \le 5\sigma$ are found, and a test-particle insertion was performed at the two points which complete an equilateral triangle. A histogram was used to measure $P^{(1)}(r,r,r)$ based on binning each measurement with a bin width of $0.01\sigma$. Note that the measurement was performed with a resolution $5\times$ greater than in the two-insertion and three-insertion cases, due to the improved statistics provided by one-particle insertion method. In the isosceles cases, where one or two of the side lengths are fixed at, e.g.\ $r_{\mathrm{m}} = 1.105$ (for $\rho\sigma^2 \approx 0.41$), pairs of particles were found with a separation $r$ between, e.g.\ $1.100$ and $1.110$, and four [Fig.\ 2(c)] or two [Fig.\ 2(d)] test-particle insertions were attempted around each pair to measure $P^{(1)}$. Note that in Fig.\ 2(c), the second side length corresponding to $r_{\mathrm{m}}$ was taken to be consistent with the exact separation of the pair, rather than as, e.g.\ $1.105$. In both cases, the triangle inequality limits the possible values of $s$ to $s \le 2 r_{\mathrm{m}}$ and $s \ge r_{\mathrm{m}}/2$, respectively. In all three cases a separate measurement of $P$ was performed using $10^4$ test particles, and this was used with the value for $g^{(2)}$ corresponding to the separation of the existing pair to calculate $g^{(3)}$ in accordance with Eq.\ (10c). Note that the one-particle insertion result for $g^{(2)}(r)$ was used for this. While pairs of particles with side lengths close to $s$ can also be considered, we do not include them in this analysis. (Note that $r_{\mathrm{m}}=1.085$ at $\rho\sigma^2 \approx 0.60$, and pairs with separation $r$ between $1.080$ and $1.090$ were found. For $\rho\sigma^2 \approx 0.77$, $r_{\mathrm{m}}=1.065$ and pairs with separation $r$ between $1.060$ and $1.070$ were found.)

\bibliography{many_body_distribution}